# Structural characterisation of high-mobility $Cd_3As_2$ films crystallised on $SrTiO_3$

Yusuke Nakazawa[1], Masaki Uchida[1,*], Shinichi Nishihaya[1], Markus Kriener[2], Yusuke Kozuka[1], Yasujiro Taguchi[2], and Masashi Kawasaki[1,2]

[1]Department of Applied Physics and Quantum-Phase Electronics Center (QPEC), the University of Tokyo, Tokyo, 113-8656, Japan
[2]RIKEN Center for Emergent Matter Science (CEMS), Wako, 351-0198, Japan
*uchida@ap.t.u-tokyo.ac.jp

## ABSTRACT

$Cd_3As_2$ has long been known as a high-mobility semiconductor. The recent finding of a topological semimetal state in this compound has demanded growth of epitaxial films with high crystallinity and controlled thickness. Here we report the structural characterisation of $Cd_3As_2$ films grown on $SrTiO_3$ substrates by solid-phase epitaxy at high temperatures up to 600 °C by employing optimised capping layers and substrates. The As triangular lattice is epitaxially stacked on the Ti square lattice of the (001) $SrTiO_3$ substrate, producing (112)-oriented $Cd_3As_2$ films exhibiting high crystallinity with a rocking-curve width of 0.02° and a high electron mobility exceeding 30,000 $cm^2$/Vs. The systematic characterisation of films annealed at various temperatures allowed us to identify two-step crystallisation processes in which out-of-plane and subsequently in-plane directions occur with increasing annealing temperature. Our findings on the high-temperature crystallisation process of $Cd_3As_2$ enable a unique approach for fabricating high-quality $Cd_3As_2$ films and elucidating quantum transport by back gating through the $SrTiO_3$ substrate.

**Introduction**

Since the topological Dirac semimetal state in $Cd_3As_2$ has been theoretically predicted and experimentally verified,[1–3] a variety of its syntheses, such as by melt growth,[1–5] Cd flux growth,[6–9] chemical vapour transport,[10–13] and chemical vapour deposition[14] have been reported so far. On the other hand, $Cd_3As_2$ has long been known as a high-mobility semiconductor,[15] and growth of rather thick films, mainly by thermal evapouration and pulsed-laser evapouration, has been reported since the 1970s.[16–22] Following the discovery of the Dirac semimetal state, molecular beam epitaxy (MBE) has been employed to investigate quantum transport phenomena in $Cd_3As_2$ thin films.[23–25] However, crystallinity and flatness of these films are still limited, mainly due to the low temperature ($\sim 200$ °C) growth necessary for avoiding the revapourisation of $Cd_3As_2$ itself. To overcome this issue, we have recently developed a high-temperature annealing method which allows to fabricate very thin (12 nm $\sim$ 20 nm) $Cd_3As_2$ films which exhibit a clear quantum Hall effect with zero resistance.[26] Here we report the detailed structural characterisation of rather thick three-dimensional (100 nm) $Cd_3As_2$ films annealed at various temperatures and elucidate the evolution of their epitaxial crystallisation and its relation to transport properties.

## Results

High-quality $Cd_3As_2$ epitaxial films are fabricated on (001) $SrTiO_3$ substrates by high temperature annealing. The $SrTiO_3$ substrates are etched using buffered hydrofluoric acid by a supplier (SHINKOSHA Co. Ltd). As shown in the x-ray diffraction (XRD) pattern later, the $Cd_3As_2$ film is amorphous just after the deposition at room temperature, necessitating an annealing process for crystallisation. $Si_3N_4$ and $TiO_2$ were deposited in-situ on the $Cd_3As_2$ film as capping layers, which prevent re-evapouration of the $Cd_3As_2$ film during the high-temperature annealing. This combination of the capping layers is chosen due to the chemical inertness of $TiO_2$ against $Cd_3As_2$ even if there is direct contact between two and the mechanical toughness of $Si_3N_4$ covering the whole film. The optimised capping layers enable annealing at temperatures as high as 600 °C, where the vapour pressure of $Cd_3As_2$ becomes increasingly high ( $\sim 10$ Torr at 600 °C[27]). Detailed growth conditions are described in Methods section.

To better understand the epitaxial relation between $Cd_3As_2$ and $SrTiO_3$, their lattice structures are presented in Fig. 1. $Cd_3As_2$ forms a solid phase having a cubic Cd-deficient antifluorite ($Cd_4As_2$) structure below 715 °C, and gets successively distorted to form $\sqrt{2} \times \sqrt{2} \times 2$ and $2 \times 2 \times 4$ superstructures below 600 °C and 475 °C, respectively, accompanied with ordered displacements of Cd atoms.[28] In this $2 \times 2 \times 4$ $Cd_3As_2$, a triangular lattice is formed on the (112) lattice plane, which corresponds to the (111) lattice plane of the high-temperature cubic antifluorite structure. The crystal structure of $SrTiO_3$ is perovskite type with a square lattice on the (001) plane. The green hexagons in Fig. 1(a) depict the in-plane epitaxial relation between the (112) $Cd_3As_2$ plane and the (001) $SrTiO_3$ plane, realizing epitaxial growth of the $Cd_3As_2$ film. The length of the perpendicular line in the As triangular lattice is 3.88 Å (white arrow in the left panel), which is very close to the lattice constant of 3.91 Å in the (001) $SrTiO_3$ plane (white arrow in the right panel). Consequently, there are two distinct stacking patterns of the (112) $Cd_3As_2$ plane on the (001) $SrTiO_3$ plane, where the [11$\bar{1}$] in-plane $Cd_3As_2$ axis is along either the [100] or [010] direction in $SrTiO_3$, as shown in the right panel of Fig. 1(a). Figures 1(c) and (d) show cross-section high-angle annular dark-field scanning transmission electron microscopy (HAADF-STEM) image along with a depth profile of each element obtained by energy dispersive x-ray spectroscopy (EDX) for the $Cd_3As_2$ film annealed at the highest temperature of 600 °C, and a schematic sketch is shown in Fig. 1(e) indicating a possible atomic structure. Incidentally, Sr EDX counts in the depth profile are suppressed in the interfacial layers as compared to the Ti and O counts, indicating that a few $TiO_2$ layers are formed at the heterointerface. Such surface termination with a few $TiO_2$ layers is known to usually occur when $SrTiO_3$ substrates are annealed at such high temperature.[29,30]

XRD $\theta$-$2\theta$ scans and rocking curves of the film peaks are summarised in Fig. 2 for the $Cd_3As_2$ films annealed at various temperatures. The diffraction pattern of the as-grown film shows no film peaks, indicating that the film is amorphous. For the film annealed at 500 °C, weak film peaks assigned to the (224) and (336) plane reflections are observed in the $\theta$-$2\theta$ scan, while a full width at half maximum (FWHM) of the rocking curve for the (224) peak is very broad (9.9°). The $\theta$-$2\theta$ scan for the film annealed at 550 °C shows much stronger peaks originating from the {112} lattice planes, and the FWHM of the rocking curve also became much sharper (0.027°). On the other hand, impurity peaks ascribed to As and $CdAs_2$ phases are detected when annealing at this temperature. By increasing the annealing temperature up to 600 °C, a (112)-oriented single-phase $Cd_3As_2$ film is obtained, as shown in Fig. 2(f). The FWHM of the rocking curve is very sharp (0.023°), which is nearly one-fourth of values reported for single-crystalline bulk samples.[7] In both Figs. 2(e) and (f), a rather broad background with weak intensity can be seen (note the logarithmic scale). In case of thinner $Cd_3As_2$ films ($\lesssim$ 20 nm), this background is not discernible. Presumably, some disorder is present for thicker films away from the $Cd_3As_2$ / $SrTiO_3$ interface.

In-plane reciprocal space mappings and $\phi$-scans for the (4$\bar{4}$0) $Cd_3As_2$ peak are presented in Figs. 2(i)-(n). For the $Cd_3As_2$ films annealed at 500 °C and 550 °C, Debye-Scherrer ring patterns are observed in the reciprocal space mappings and no sharp peaks are confirmed in the $\phi$-scans. These results indicate that the $Cd_3As_2$ films annealed below 550 °C are not oriented along the in-plane directions, while they are crystallised with the [112] out-of-plane orientation as confirmed in Figs. 2(b) and (d). In contrast, the reciprocal space mapping for the film annealed at 600 °C shows a clear peak from the $(4\bar{4}0)$ reflection plane, which exhibits a six-fold symmetry in the $\phi$-scan. In addition, other six peaks with much weaker intensity are observed between the main peaks in the $\phi$-scan. The existence of these two sets of peaks with contrasting intensities indicates that there are two types of stacking patterns originating from the epitaxial relation as shown in Fig. 1(a). One of them becomes dominant through the high-temperature annealing, which is possibly due to a miscut direction of the $SrTiO_3$ substrates.

To investigate the in-plane orientation in more detail, a planer TEM image is taken for the film annealed at the highest temperature of 600 °C. Figure 3(a) shows the existence of domains as large as $> 10$ $\mu m^2$. A higher-resolution magnified image of a tri-sectional domain boundary is presented in Fig. 3(b). As shown in the insets, the obtained electron diffraction patterns agree well with the one simulated for the incident beam direction along $Cd_3As_2$ [112]. All these three diffraction patterns correspond to the major stacking pattern of the (112) $Cd_3As_2$ plane depicted as a dark green hexagon in Fig. 1(a), being consistent with the major peaks appearing in in-plane $\phi$-scan for the $Cd_3As_2$ $(4\bar{4}0)$ plane shown in Fig. 2(n). Although these three domains are in the square epitaxial relation, the in-plane orientation exhibits a small variance of several degrees in the tilting angle. This variance of the in-plane orientation also explains the broad peaks observed in the $\phi$-scan, indicating that the small in-plane misorientation is the origin of the domain formation.

## Discussion

Annealing temperature dependences of the film crystallinity and transport properties are summarised in Fig. 4. As detailed in Fig. 2, the FWHM of the rocking curve for the out-of-plane (224) peak sharply drops from 500 °C to 550 °C (red in Fig. 4(a)), whereas annealing at a temperature as high as 600 °C is needed to promote the in-plane orientation alignment (green in Fig. 4(a)). The considerable increase in conductivity occurs between 550 °C and 600 °C, suggesting that the in-plane alignment plays an important role.

The carrier density ($n$, orange) and electron mobility ($\mu$, blue) of the $Cd_3As_2$ films for each annealing temperature are deduced from the Hall measurement, as plotted in Fig. 4(b). The carrier density agrees well with one deduced from the SdH oscillations. Before annealing, $n$ is $7 \times 10^{18}$ $cm^{-3}$ and it is significantly reduced to $9 \times 10^{17}$ $cm^{-3}$ by annealing at temperatures up to 600 °C. This reduction in $n$ may be explained as follows. As shown in the XRD pattern for the film annealed at 550 °C (Fig. 2(d)), arsenic-rich impurity phases (As and $CdAs_2$) are confirmed, presumably resulting in arsenic deficiency in the $Cd_3As_2$ phase. Since $Cd_3As_2$ is known to be naturally $n$-type due to As deficiency, $n$ of the $Cd_3As_2$ films heated at 600 °C is reduced by the annealing-induced chemical reaction towards a more stoichiometric phase. The mobility $\mu$ reaches $3.4 \times 10^4$ $cm^2/Vs$ after annealing at 600 °C, while it is $1.5 \times 10^3$ $cm^2/Vs$ before annealing. The increase in the mobility up to the annealing temperature of 550 °C is mainly attributed to the reduction of the carrier density as seen above. The further increase of the mobility from 550 °C to 600 °C is due to the reduction of grain boundaries from in-plane random to epitaxially locked orientations. Fig. 4(c) shows longitudinal magnetoresistances of these samples measured with applying the magnetic field perpendicular to the film plane. They are normalised to the zero-field results. With increasing annealing temperature, Shubnikov-de Haas oscillations and quadratic positive magnetoresistance become more pronounced, demonstrating that quantum transport is achievable in films of high-crystallinity and high-mobility.

In summary, we have performed a detailed structural characterisation of high-crystallinity $Cd_3As_2$ films fabricated by high-temperature solid-phase epitaxy. From the systematic characterisation of the films annealed at various high temperatures, successive crystallisation processes take place about the out-of-plane and in-plane orientations. The electron mobility is strikingly enhanced by the epitaxial crystallisation on the square-lattice and the effective reduction of arsenic deficiency. The mobility is expected to be further enhanced by reducing the number of domain boundaries and point defects. Our systematic characterisation of high-quality $Cd_3As_2$ films grown on dielectric oxides provides the foundation to prepare higher-quality $Cd_3As_2$ films by reducing the in-plane domains and carrier densities as well as to investigate quantum transport phenomena by back gating and chemical substitution.

## Methods

Prior to this study, we have screened various materials and tested them for the usefulness as capping layer and substrate. The best results are achieved when using an optimised combination of $TiO_2/Si_3N_4$ and (001) $SrTiO_3$, respectively.[26] For preparing a $Cd_3As_2$ target, 6N5 Cd and 7N5 As shots were mixed at the stoichiometric ratio and heated at 950 °C for 48 hours in a vacuum-sealed silica tube. After heating the mixture, it was grinded and pelletised and then it was sintered at 250 °C for 30 hours in a vacuum-sealed tube. The $Cd_3As_2$ target was ablated using KrF excimer laser at room temperature and a base pressure of about $2 \times 10^{-7}$ Torr. The laser fluence and frequency were set to 0.6 J/cm$^2$ and 10 Hz, respectively. Subsequently, 30 nm $TiO_2$ and 100 nm $Si_3N_4$ capping layers were deposited by ablating their targets with a laser fluence of 4.0 J/cm$^2$ and a frequency of 20 Hz. After all the layers were deposited, the sample was cut into pieces and each pieces was annealed in air at a temperature of 500 °C, 550 °C, and 600 °C for 5 minutes in a rapid thermal annealing system. Annealing at higher temperatures resulted in cracking of the capping layers due to the high vapour pressure of $Cd_3As_2$. Thicknesses of the respective layers were confirmed from Laue oscillations in the XRD $\theta$-$2\theta$ scan.

## Acknowledgements

This work was supported by JST CREST Grant No. JPMJCR16F1, Japan and by Grant-in-Aids for Scientific Research on Innovative Areas "Topological Materials Science" No. JP16H00980 and Scientific Research (C) No. JP15K05140, from MEXT, Japan.

## Author contributions statement

Y.N., M.U., and M.Kawasaki designed the experiments. Y.N., M.U., and S.N. synthesised the bulk target with M.Kriener and performed film growth and transport measurements. Y.N. and M.U. analysed the data and wrote the manuscript with contributions from all the authors. Y.K., Y.T., and M.Kawasaki jointly discussed the results. M.U. and M.Kawasaki conceived the project. All authors reviewed the manuscript.

## Additional information

The authors declare that they have no competing interests.

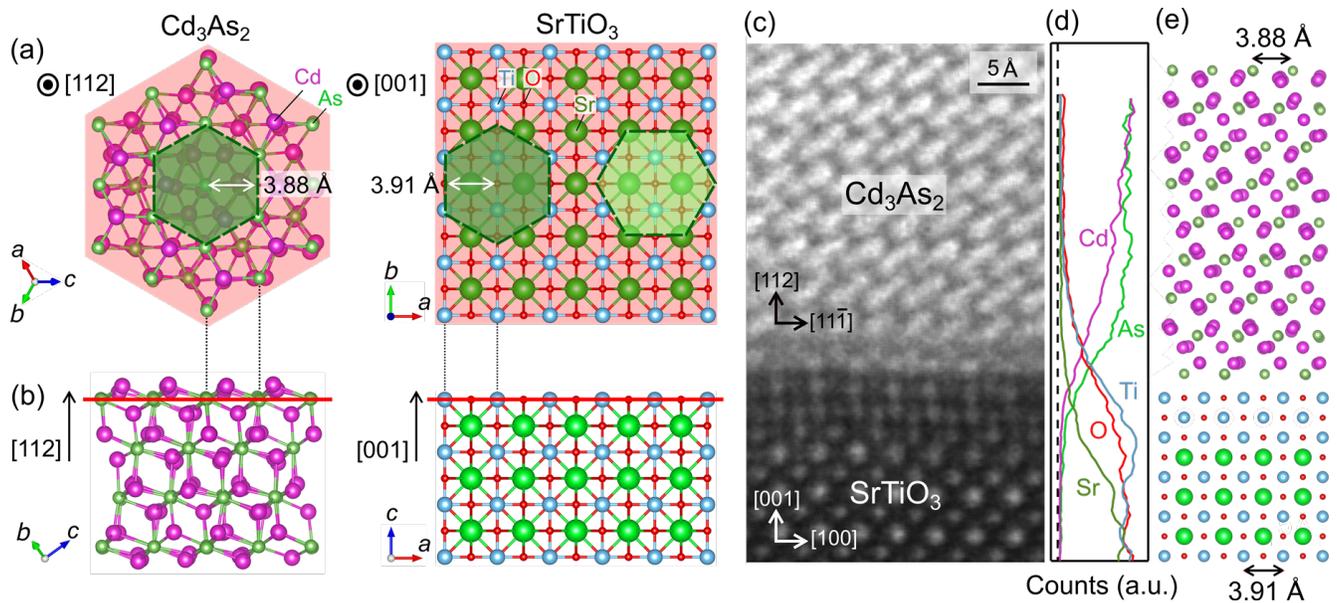

**Figure 1. Epitaxial relation between $Cd_3As_2$ and $SrTiO_3$.** (a) Top and (b) side views of crystal structures, representing epitaxial [112] $Cd_3As_2$ / [001] $SrTiO_3$ relation. The green hexagons in (a) represent the As triangular lattice in $Cd_3As_2$. There are two possible in-plane alignments on the Ti square lattice (right). (c) HAADF-STEM image showing atomic arrangement at the heterointerface between $Cd_3As_2$ film and $SrTiO_3$ substrate. (d) Depth profile of Cd, As, Sr, Ti, and O, obtained by integrating EDX counts along the horizontal direction in (c). Incidentally, Sr EDX counts in the depth profile are certainly suppressed in the interfacial layers as compared to the respective Ti and O results, indicating that a few $TiO_2$ layers are formed by high-temperature annealing.[29,30] The STEM and EDX measurements were performed using a reduced acceleration voltage of 80 kV to reduce the damage at the $Cd_3As_2$ interface, resulting in lower resolution compared to images of the central region[26] taken with an acceleration voltage of 200 kV. The damage is inevitable to some extent, causing deviation from the Cd/As stoichiometric composition at the interface. (e) Epitaxial relation of the (112) oriented $Cd_3As_2$ film on the (001) $SrTiO_3$ substrate. Projected lattice distances of the triangular-lattice As atoms and the square-lattice Ti atoms are almost the same.

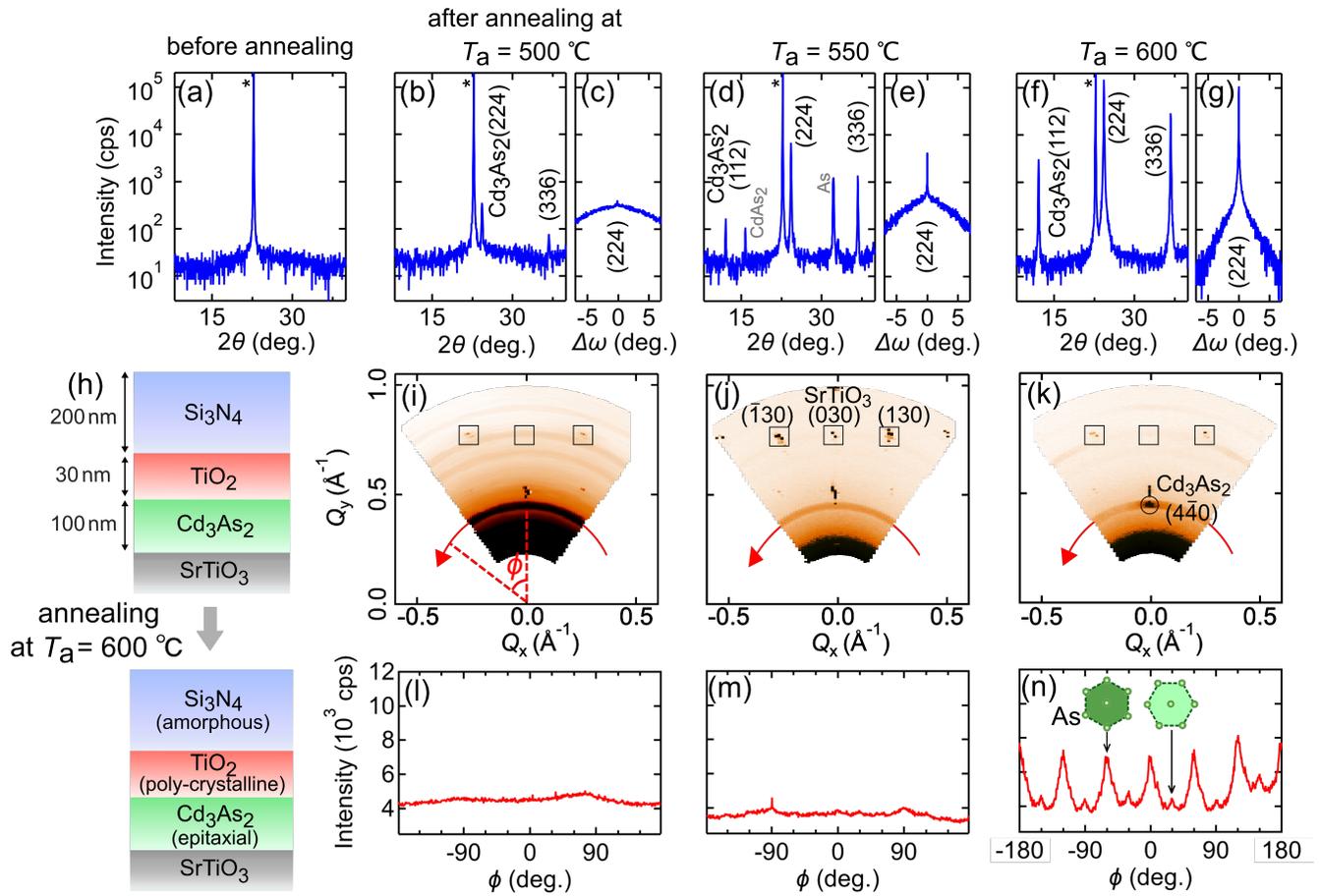

**Figure 2. XRD characterisation of the Cd$_3$As$_2$ films annealed at high temperatures.** XRD $\theta$-$2\theta$ scans and corresponding rocking curves of the (224) Cd$_3$As$_2$ film peaks (a) before annealing and (b),(c) after annealing at 500 °C, (d),(e) 550 °C, and (f),(g) 600 °C. (h) Sample structure and changes of the crystalline nature in each layer due to the annealing. (i)-(k) In-plane reciprocal space mappings and (l)-(n) $\phi$-scans along the red curves in the reciprocal space mappings. The $\phi$-scan pattern shown in (n) represents two sets of 6-fold peaks. Major and minor in-plane stacking patterns of the As triangular lattice are denoted by dark and light green hexagons as shown in the right panel of Fig. 1(a).

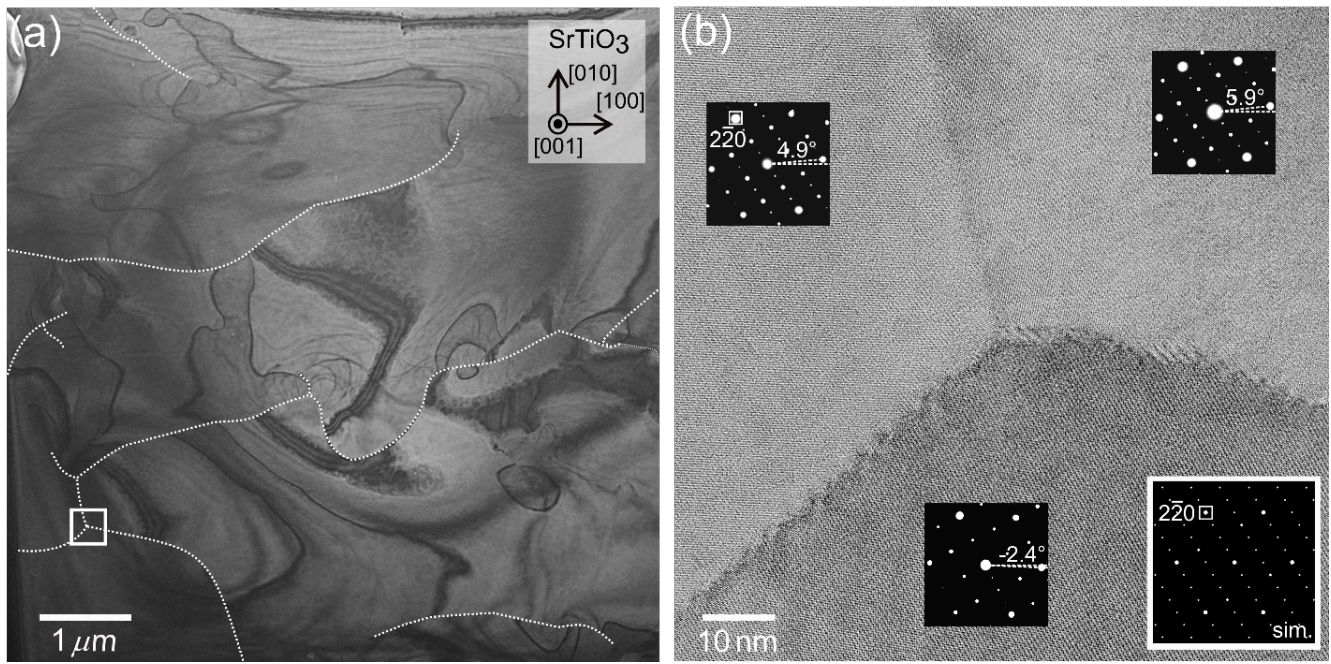

**Figure 3. Plane-view TEM characterisation of the Cd$_3$As$_2$ film.** (a) Planar TEM picture of the film annealed at the highest temperature of 600 °C. Dotted curves are overlaid on discernible domain boundaries. Crystal axes of the SrTiO$_3$ substrate are shown in the inset for reference. (b) Higher-resolution magnified image focusing on a tri-sectional grain boundary in the boxed area in (a). Electron diffraction patterns taken in the respective domains are shown in the insets. The brightest spot at the center is the zeroth diffraction spot. The diffraction patterns in each grain are slightly tilted as compared to the simulated one (right bottom), where the in-plane crystallographic axes of the Cd$_3$As$_2$ film are assumed to be completely aligned with the respective SrTiO$_3$ axes.

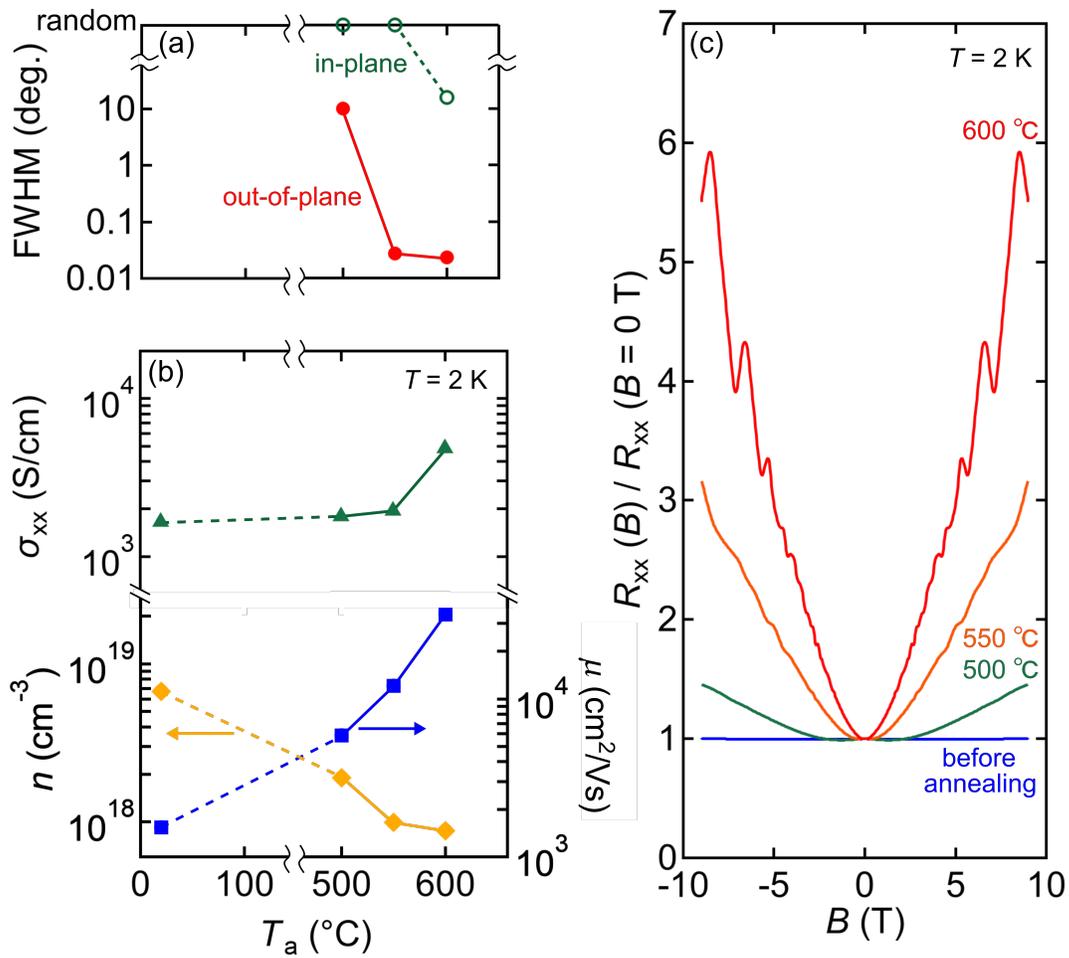

**Figure 4. Evolution of crystallinity and transport properties.** (a) Annealing temperature dependence of the rocking curve width for the out-of-plane (224) $Cd_3As_2$ film peak and the in-plane $(4\bar{4}0)$ film peak. (b) Same plots for longitudinal conductivity $\sigma_{xx}$, carrier density $n$, and the electron mobility $\mu$ at 2 K. (c) Normalised longitudinal magnetoresistance.